\documentclass{INTERSPEECH2023}
\usepackage{cite}

\interspeechcameraready

\title{Creating Personalized Synthetic Voices from Post-Glossectomy Speech with Guided Diffusion Models}
\name{Yusheng Tian, Guangyan Zhang, Tan Lee}
\address{
  Department of Electronic Engineering, The Chinese University of Hong Kong, Hong Kong SAR}
\email{\{ystian0617, gyzhang\}@link.cuhk.edu.hk, tanlee@ee.cuhk.edu.hk}

\begin{document}

\maketitle
 
\begin{abstract}
This paper is about developing personalized speech synthesis systems with recordings of mildly impaired speech. In particular, we consider consonant and vowel alterations resulted from partial glossectomy, the surgical removal of part of the tongue. The aim is to restore articulation in the synthesized speech and maximally preserve the target speaker’s individuality. We propose to tackle the problem with guided diffusion models. Specifically, a diffusion-based speech synthesis model is trained on original recordings, to capture and preserve the target speaker’s original articulation style. When using the model for inference, a separately trained phone classifier will guide the synthesis process towards proper articulation. Objective and subjective evaluation results show that the proposed method substantially improves articulation in the synthesized speech over original recordings, and preserves more of the target speaker’s individuality than a voice conversion baseline.
\end{abstract}

\section{Introduction}
People diagnosed with oral cancer may need to have the entire voice box removed by surgery, losing the ability to speak thereafter. One way to save their voices is to collect audio recordings of their speech before the scheduled operation date and create personalized text-to-speech (TTS) models from those speech data. The TTS systems would allow these individuals to communicate with other people using their own voices. Ideally the speech used to create TTS models should carry accurate and clear pronunciation such that highly intelligible speech can be generated. In reality, some of the patients already suffer speech impairment at the time of recording. For example, tongue cancer, one of the most common sites of oral cancer, is often treated with surgery. People who underwent tongue surgery typically show consonant and vowel alterations \cite{acher2014speech,girod2020rehabilitation}. We encountered one such case recently. A young female Cantonese speaker approached us, expressing the hope of saving her voice through speech synthesis technology. She received partial glossectomy six years ago, and about 3/4 of her tongue was removed surgically. This has resulted in significant difficulties in articulating accurately due to the defect in the tongue. Consonant and vowel alterations are reflected noticeably in this young lady’s speech recordings. This application context leads to the task committed in the present study: developing a personalized TTS system from post-glossectomy speech. Our goal is to restore articulation in the synthesized speech and maximally preserve the target speaker’s individuality.

The problem of voice reconstruction from impaired speech was tackled in a few previous studies. They share the same idea of substituting impaired speech segments by normal ones. The Voicebank project \cite{yamagishi2012speech,veaux2012using,creer2013building} applied HMM-based speech synthesis techniques to create personalized synthetic voices for people with speech disorder. Synthetic voices were repaired by substituting selected acoustic feature parameters with those of an average healthy voice. In \cite{nanzaka2018hybrid, matsubara2021high}, neural speech synthesis systems were developed for individuals with dysarthria. A two-step process was developed: first train a neural TTS model on recordings from a healthy speaker, then perform voice conversion on the synthesized speech from the healthy speaker to the impaired speaker.

The above substitution-based approaches are considered suboptimal for the task we are considering, where the major symptoms of speech disorder is alteration of certain phones. On the one hand, the articulation style is part of an individual’s personality. Preserving only the voice timbre as in \cite{nanzaka2018hybrid, matsubara2021high} would completely discard the target speaker’s articulation style, including both bad and good aspects, leading to undesirable loss of individuality. On the other hand, although HMM-based systems can allow acoustic feature substitution on selected phones \cite{veaux2012using}, the average voice model may not provide a replacement close enough to the target speaker's voice. 

We propose to restore articulation in the synthesized speech with guided diffusion models. Specifically, a diffusion-based TTS model\cite{popov2021grad} is trained on the target speaker’s original recordings (with impaired articulation on certain speech sounds), and hence capture and preserve the original articulation style. When using the trained model for inference, the synthesis process is guided by the gradient from a separately trained phone classifier. The output of the phone classifier indicates how precise the articulation is on a continuous scale, which facilitates finer adjustment of articulation than simple substitution.

The proposed system design is inspired by GuidedTTS ~\cite{kim2022guided}, which employs a phone classifier to guide an unconditional diffusion model for speech generation. However, the motivation of using classifier guidance in GuidedTTS is to develop TTS systems with untranscribed speech data. The phone classifier guidance in the present study is in the spirit more similar to shallow-fusion biasing in end-to-end automatic speech recognition (ASR) ~\cite{shallowfusion1williams2018contextual, shallowfusion2zhao2019shallow}, in which an external language model steers the recognition process towards a particular context at each decoding step. In the proposed system, the external phone classifier guides the synthesis process towards correct pronunciation at each reverse diffusion step.

In the next section, we outline key concepts in diffusion models that are related to diffusion-based speech synthesis. Section 3 describes the proposed system of guided speech synthesis. Section 4 and 5 present the experimental results on the aforementioned real patient case. Section 6 concludes and discusses the limitations of the proposed approach.

\section{Background}
Diffusion models are a family of probabilistic generative models. The modelling involves a forward process that progressively contaminates the data with random noise, and a reverse process that generates data from random noise. The forward and backward process were originally formulated as Markov chains \cite{sohl2015deep,song2019generative,ho2020denoising}. In \cite{song2020score}, the discrete-time Markov process is generalized to a continuous-time stochastic differential equation (SDE). Specifically, the forward process is defined by the following equation:
\begin{align}
\mathrm{d}\boldsymbol{x}_t&=\boldsymbol{f}(\boldsymbol{x}_t, t)\mathrm{d}t+g(t)\mathrm{d}\boldsymbol{w}_t\ ,
\label{forwardSDE}
\end{align}
where \(t\sim\mathcal{U}(0,T)\); \(\boldsymbol{w}_t\) is a standard Brown motion; \(\boldsymbol{f}(\cdot)\) and \(g(\cdot)\) are the so-called drift and diffusion coefficient. These two coefficients are constructed such that as \(t\) grows from \(0\) to \(T\), the probability distribution of \(\boldsymbol{x}_t\) would evolve from the original data distribution to a tractable prior, typically a Gaussian distribution with fixed mean and variance.

The corresponding reverse process also forms a SDE as stated in \cite{anderson1982reverse}:
\begin{align}
\mathrm{d}\boldsymbol{x}_t&=\left[\boldsymbol{f}(\boldsymbol{x}_t, t)-g^2(t)\nabla_{\boldsymbol{x}_t}\log P(\boldsymbol{x}_t) \right]\mathrm{d}t+g(t)\mathrm{d}\bar{\boldsymbol{w}_t}
\label{equation:reverseSDE}
\end{align}
where \(\bar{\boldsymbol{w}}_t\) is a standard Brown motion running backward in time. The core part of a diffusion model is to train a neural network \({S}_{\boldsymbol{\theta}}\) to estimate the value of \(\nabla_{\boldsymbol{x}_t}\log P(\boldsymbol{x}_t)\), a.k.a., the score. Once the score is known for all \(t\), we can sample data by solving the reverse SDE using numerical solvers.

It was proved in \cite{song2020score} that the reverse SDE is associated with an ordinary differential equation (ODE) as follows:
\begin{align}
\mathrm{d}\boldsymbol{x}_t&=\left[\boldsymbol{f}(\boldsymbol{x}_t, t)-\frac{1}{2}g^2(t)\nabla_{\boldsymbol{x}_t}\log P(\boldsymbol{x}_t) \right]\mathrm{d}t\ ,
\label{equation:reverseODE}
\end{align}
which shares the same marginal distribution for all \(t\). It is empirically shown in \cite{popov2021grad} that inference with the ODE formulation requires fewer sampling steps. In the remainder of this paper, we use the ODE formulation to model the reverse process.



\section{Approach}

\subsection{Diffusion-based TTS}
We follow GradTTS \cite{popov2021grad} and use diffusion models to generate Mel-Spectrograms conditioned on the input text and speaker labels. The forward process SDE is defined as:
\begin{align}
    \mathrm{d}\boldsymbol{x}_t &=\frac{1}{2}\left(\boldsymbol{\mu}-\boldsymbol{x}_t\right)\beta_t\mathrm{d}t+\sqrt{\beta_t}d\boldsymbol{w}_t\ ,
    \label{equation:GradTTS_forward}
\end{align}
where \(t\sim\mathcal{U}(0,1)\), \(\beta_t =\beta_0 + \left(\beta_1 - \beta_0 \right)t\) is a predefined linear noise scale, and \(\boldsymbol{\mu}\) is an average Mel-spectrogram corresponding to the input phone sequence. 
%
%
One important result derived from (\ref{equation:GradTTS_forward}) is the conditional distribution of \(\boldsymbol{x}_t\) given \(\boldsymbol{x}_0\):
\begin{align}
    P(\boldsymbol{x}_t|\boldsymbol{x}_0) = \mathcal{N}(\boldsymbol{\rho}(\boldsymbol{x}_0, t), \sigma_t^2\boldsymbol{I})\ ,
    \label{equation:conditional}
\end{align}
where \(\boldsymbol{\rho}(\boldsymbol{x}_0, t)=(1-e^{-\frac{1}{2}\int_0^t\beta_s \mathrm{d}s})\boldsymbol{\mu} + e^{-\frac{1}{2}\int_0^t\beta_s \mathrm{d}s}\boldsymbol{x}_0\), and \(\sigma_t^2=1-e^{-\int_0^t \beta_s \mathrm{d}s}\). If \(\boldsymbol{x}_0\) is known, we can then draw samples of \(\boldsymbol{x}_t\) using the reparameterization trick:
\begin{align}
    \boldsymbol{x}_t = \boldsymbol{\rho}(\boldsymbol{x}_0,t) + \sigma_t\boldsymbol{\epsilon}_t\ ,\boldsymbol{\epsilon}_t \sim \mathcal{N}(\mathbf{0}, \boldsymbol{I})\ .
    \label{equation:reparameterization}
\end{align}
The reverse time ODE is given by:
\begin{align}
    \mathrm{d}\boldsymbol{x}_t &= \frac{1}{2}\beta_t \left[\boldsymbol{\mu}-\boldsymbol{x}_t - \nabla_{\boldsymbol{x}_t}\log P(\boldsymbol{x}_t|\boldsymbol{\mu}, s)\right]\mathrm{d}t\ ,
    \label{equation:GradTTSreverse}
\end{align}
where \(s\) stands for the speaker label. Note that that unlike the unconditional reverse process given by (\ref{equation:reverseODE}), the reverse process in the context of TTS is conditioned on the input text and the speaker label.

The neural network \(S_{\boldsymbol{\theta}}\) is trained to predict the conditional score function \(\nabla_{\boldsymbol{x}_t}\log P(\boldsymbol{x}_t|\boldsymbol{\mu}, s)\) using a weighted L2 loss:
\begin{align}
    \mathcal{L}(\boldsymbol{\theta})=\mathbb{E}_{t} \sigma_t^2{\mathbb{E}}_{\boldsymbol{x}_0} {\mathbb{E}}_{\boldsymbol{\epsilon}_t}\Vert{S}_{\boldsymbol{\theta}}(\boldsymbol{x}_t, t, \boldsymbol{\mu}, s)+\sigma_t^{-1}\boldsymbol{\epsilon}_t\Vert_2^2\ ,
    \label{equation:loss}
\end{align}
where we have made use of the following results:
\begin{align}
    P\left(\boldsymbol{x}_t|\boldsymbol{x}_0, \boldsymbol{\mu}, s\right)=P\left(\boldsymbol{x}_t|\boldsymbol{x}_0\right)=\mathcal{N}(\boldsymbol{\rho}(\boldsymbol{x}_0, t), \sigma_t^2\boldsymbol{I})\ ,
    \label{equation:fact1}
\end{align}
\vspace{-2.2em}
\begin{align}
    \nabla_{\boldsymbol{x}_t}\log P(\boldsymbol{x}_t|\boldsymbol{x}_0)=-\sigma_t^{-1}\boldsymbol{\epsilon}_t\ .
    \label{equation:fact2}
\end{align}

\subsection{Guided synthesis process}
\begin{figure*}[t]
  \centering
  \includegraphics[width=0.75\linewidth]{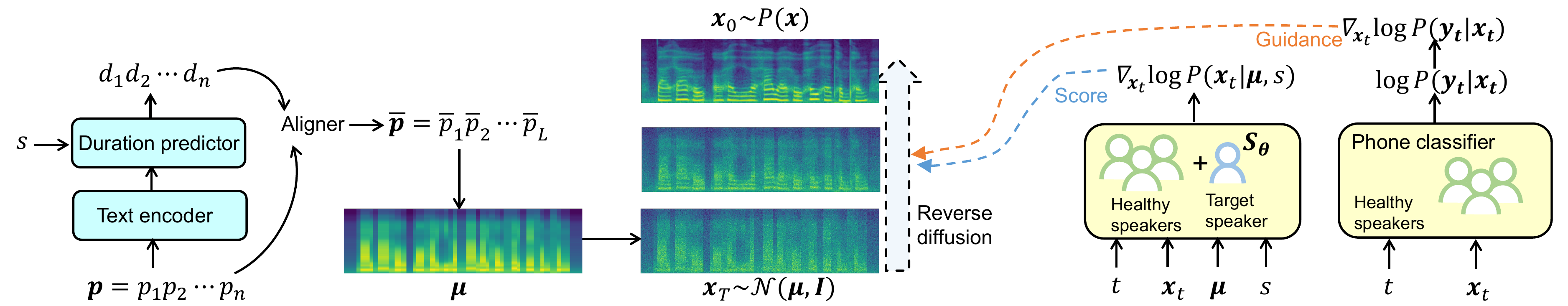}
  \caption{Schematic diagram of the proposed guided speech synthesis process.}
  \vspace{-1.7em}
  \label{fig:system_design}
\end{figure*}

Once the diffusion-based acoustic model is trained, the target speaker’s articulation style will be embedded into the generated Mel-Spectrogram through the speaker condition \(s\). The synthetic voice will inevitably preserve the problematic articulation pattern. In order to improve articulation in the synthesized speech, another condition \(\boldsymbol{y}_t\) is provided to the score estimator. \(\boldsymbol{y}_t\) stands for the phone label sequence of \(\boldsymbol{x}_t\) predicted by an external frame-level phone classifier, i.e., what the external phone classifier thinks the synthetic voice is saying. Now the reverse time ODE is augmented as
\begin{align}
    \mathrm{d}\boldsymbol{x}_t &= \frac{1}{2}\beta_t(\boldsymbol{\mu}-\boldsymbol{x}_t)\mathrm{d}t \\ \nonumber
    &- \frac{1}{2}\beta_t\left[\nabla_{\boldsymbol{x}_t}\log P(\boldsymbol{x}_t|\boldsymbol{\mu}, s)+\nabla_{\boldsymbol{x}_t}\log P(\boldsymbol{y}_t|\boldsymbol{x}_t)\right]\mathrm{d}t\ ,
\end{align}
where we have assumed that the phone classifier is speaker-independent, i.e. \(\log P(\boldsymbol{y}_t|\boldsymbol{x}_t,\boldsymbol{\mu}, s)=\log P(\boldsymbol{y}_t|\boldsymbol{x}_t)\).

To see why incorporating this external phone classifier can improve the articulation in the synthetic voice, consider the following scenarios. First, suppose the articulation is not good, i.e. a low value of \(P(\boldsymbol{y}_t|\boldsymbol{x}_t)\). Thus the additional term \(\nabla_{\boldsymbol{x}_t}\log P(\boldsymbol{y}_t|\boldsymbol{x}_t)\) will bias the synthesis process towards a direction that increases \(\log P(\boldsymbol{y}_t|\boldsymbol{x}_t)\). If the phone classifier is trained on speech from healthy speakers, then the articulation would be improved as \(\log P(\boldsymbol{y}_t|\boldsymbol{x}_t)\) grows. Second, suppose the articulation is good, i.e., a high log probability of \(P(\boldsymbol{y}_t|\boldsymbol{x}_t)\). In such case the additional term is close to zero and will have little impact on the synthesized speech. Consequently, the good part of the target speaker's articulation style will be preserved.

In practice, we usually scale \(\nabla_{\boldsymbol{x}_t}\log p_t(\boldsymbol{y}_t|\boldsymbol{x}_t)\) with a positive coefficient \(\gamma\) in order to obtain better sample quality\cite{kim2022guided, BeatGANdhariwal2021diffusion}. We adopt the norm-based scale\cite{kim2022guided}, which is computed as \(\gamma=\alpha\cdot\Vert\nabla_{\boldsymbol{x}_t}\log P(\boldsymbol{x}_t|\boldsymbol{\mu}, s)\Vert_2/\Vert\nabla_{\boldsymbol{x}_t}\log P(\boldsymbol{y}_t|\boldsymbol{x}_t)\Vert_2\), where \(\alpha\) is a hyperparameter to be tuned.

Assume that the prediction of phone label for each frame is independent of each other, then we have
\begin{align}
    \nabla_{\boldsymbol{x}_t}\log P(\boldsymbol{y}_t|\boldsymbol{x}_t) = \sum\limits_{i=1}^L \nabla_{\boldsymbol{x}_t} \log P(y_t^{(i)} | \boldsymbol{x}_t)\ ,
    \label{equation:frame}
\end{align}
where $L$ refers to the number of acoustic frames. This property enables flexible guidance weights on different frames. For example, we can put extra weights on frames corresponding to phones that the target speaker tends to make mistakes.

\subsection{System design}


Figure \ref{fig:system_design} gives an overview of the proposed speech synthesis system. The input phone sequence \(\boldsymbol{p}\) is fed into a text encoder. The resulted phone embeddings, as well as the speaker label \(s\), are sent to a duration predictor to estimate the frame length for each phone. The input phone sequence \(\boldsymbol{p}\) is then expanded to \(\bar{\boldsymbol{p}}\) according to the predicted duration \(\boldsymbol{d}\), similar to the length regulator mechanism in \cite{ren2019fastspeech, durian, non_att_tacotron}. 

 In GradTTS \(\boldsymbol{\mu}\) is designed to be dependent on both the input phone sequence and the speaker label, but we augment it to be conditioned only on the phone sequence, with the hope that a shared phone embedding across speakers will transfer some articulation knowledge from healthy speakers to the target speaker, and therefore alleviate the articulation disorder in the synthetic voice. Specifically, \(\boldsymbol{\mu}\) is obtained by looking up a predefined phone-to-Mel-spectrum dictionary computed over training data, similar to that in \cite{unified_vc_cloning}.

The score estimator \({S}_{\boldsymbol{\theta}}\) is trained on speech data from a multi-speaker speech corpus, plus recordings from the target speaker. In contrast, the phone classifier is trained only on recordings from healthy speakers, to ensure that it is sensitive to abnormal articulation style.

\section{Experimental Setup}
\subsection{Baseline systems for comparison}
We compare the proposed system with two TTS baselines. The first baseline, \emph{DuriTaco}, is adapted from Tacotron \cite{tacotron}, with the attention module replaced by a duration-informed aligner as in DurIAN \cite{durian}. The second baseline, \emph{DuriTaco+VC}, has exactly the same architecture as DuriTaco, but is trained on voice-converted speech: unimpaired speech from a healthy source speaker converted into the target speaker's voice via a voice conversion (VC) model. We use the recently proposed NANSY \cite{nansy} for voice conversion, as it shows strong performance in cross-lingual setting. The underlying assumption is that impaired speech can be viewed as a new language, therefore a strong cross-lingual VC model is expected to perform well on voice conversion between normal and impaired speech.

\subsection{Datasets}
We use the following four datasets for different purposes.
\begin{itemize}
    \item \textbf{CUSENT}\cite{lee2002spoken} for multi-speaker TTS pre-training, as well as for training the external phone classifier and many-to-many VC model. It is a Cantonese speech corpus of around 20 hours clean read speech from 80 speakers, sampled at 16 kHz. 
    \item \textbf{Recording\_T} for personalized TTS fine-tuning. It contains 377 utterances, giving a total of approximately 24 minutes speech from our target speaker, a young Cantonese-speaking female. The recording script is adapted from CUSENT and all Cantonese initials, finals and tones are included. The audio was recorded with a TASCAM DR-44WL at 44.1 kHz in her living place, under reasonably quiet condition.
    \item \textbf{Recording\_S} as the source speaker's data for voice conversion. It contains 377 utterances that share exactly the same content as Recordings\_T, recorded by another female Cantonese speaker with no articulation disorder. The recording was carried out in a sound-proof studio.
    \item \textbf{KingASR086} for training a CTC-based ASR model for objective evaluation. It is a commercial Cantonese speech corpus purchased from SpeechOcean\footnote{\url{https://en.speechocean.com/datacenter/recognition.html}}, which contains 80-hour read speech from 136 speakers, sampled at 44.1 kHz.
\end{itemize}
\subsection{Implementation details}
We adopted the official implementation of GradTTS \footnote{\url{https://github.com/huawei-noah/Speech-Backbones/tree/main/Grad-TTS}} for the proposed model, but augmented the text encoder to be conditioned only on the input phone sequence as mentioned earlier. For simplification, we also used forced alignment \cite{mfa}, instead of monotonic alignment search to obtain the duration labels. We used the Jasper architecture\cite{li2019jasper} for the frame-level phone classifier, but removed the convolution stride so that it could predict phone labels for each frame. The DuriTaco baseline was similar to a public implementation of a baseline DurIAN\footnote{\url{https://github.com/ivanvovk/durian-pytorch}}, except that the duration predictor and the acoustic model were separately trained. The VC model in the DuriTaco+VC baseline followed a public implementation of NANSY\footnote{\url{https://github.com/dhchoi99/NANSY}}. The CTC-based ASR model for objective evaluation was implemented following the recipe from SpeechBrain\footnote{\url{https://github.com/speechbrain/speechbrain/tree/develop/recipes/TIMIT/ASR/CTC}}.

All TTS models generate 80-dimensional log Mel-Spectrograms to drive a pre-trained HiFi-GAN vocoder\cite{hifigan}. TTS audio data were resampled to 22.05 kHz for Mel-Spectrogram computation, to be consistent with the settings of HiFi-GAN. The one-hot speaker embedding dimension is 16 in all TTS models. The coefficient \(\alpha\) for classifier guidance in the proposed model is set to 0.3. We also assigned extra guidance weight of 5.0 to several selected phones that the target speaker had difficulties to articulate. All TTS models were pre-trained on CUSENT for 700 epochs at batch size 32, then fine-tuned on the target speaker's data (original or voice-converted) for 5000 steps at batch size 16. A total of 30 sentences from target speaker's data were held out for evaluation. The number of reverse steps for the proposed diffusion-based TTS model is 25 using the DPM solver \cite{lu2022dpmsolver}.

\section{Results}
\subsection{Objective evaluation}
We resynthesized all 377 sentences of the target speaker's recordings with the proposed system and the two baselines, and used the separately trained CTC-based ASR to evaluate the quality of articulation. The Phone Error Rate (PER\%) results in Table \ref{tab:similarity} show that the proposed system GuidedDiff produces significantly better articulation than the voice in the original recordings and the DuriTaco baseline, though not as good as the VC baseline. By comparing the last two rows we can conclude that the improvement comes from the classifier guidance, rather than the diffusion model.
\begin{table}
		\caption{PER(\%) on resynthesized speech and the original recordings with the separately trained ASR model.}
		\label{tab:similarity}

		\centering
		\renewcommand{\arraystretch}{0.9}
            \small
 		\begin{tabular}{lcc}
			\toprule
			Case & Audio Source & PER\%\\
			\midrule
                \midrule
			Recording\_S & Real & 11.2\\
                Recording\_T & Real & 43.1\\
                \midrule
                DuriTaco & TTS &  37.6\\
                DuriTaco + VC & TTS & \textbf{14.9} \\
                Diffusion & TTS & 36.4 \\
                GuidedDiff & TTS & \textbf{22.1}\\
			\bottomrule
		\end{tabular}
	\end{table}
\subsection{Subjective evaluation}
\begin{figure}[t]
  \centering
  \includegraphics[width=0.9\linewidth]{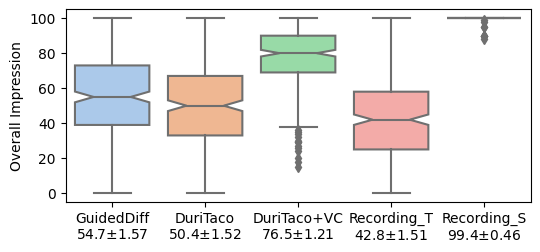}
  \caption{MUSHRA scores for overall impression. Mean value and 95\% confidence interval are reported at the bottom.}
  \label{fig:nat_mushra}
  \vspace{-1.2em}
\end{figure}

\begin{figure}[t]
  \centering
  \includegraphics[width=0.9\linewidth]{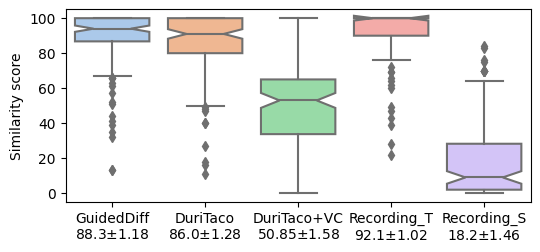}
  \caption{MUSHRA scores for speaker similarity. Mean value and 95\% confidence interval are reported at the bottom.}
  \label{fig:sim_mushra}
  \vspace{-1.2em}
\end{figure}

\begin{figure}[t]
  \centering
  \includegraphics[width=0.9\linewidth]{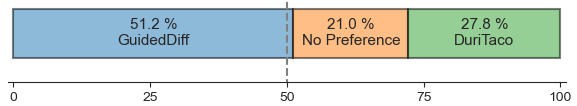}
  \caption{Preference test demonstrating that GuidedDiff produces better articulation than the DuriTaco baseline.}
  \label{fig:preference}
  \vspace{-1.7em}
\end{figure}

The subjective evaluation was conducted through a web-based listening test. It consists of three parts. The first two parts was in MUSHRA-style format (no anchor but reference was given) \cite{mushra}. Specifically, the same 30 sentences from the evaluation set were synthesized by the proposed and baseline voices. For each sentence, the three synthetic voices plus the original recordings from the target and source speaker formed five stimuli in total. In Part1, recordings from the source speaker were provided as the reference, and listeners were asked to rate the \textit{overall impression} of each stimulus on a scale from 0 to 100. We avoided the term \textit{naturalness} as it might be confusing whether disordered speech from a natural person count as natural. Part2 was similar to Part1, except that this time recordings from the target speaker were provided as the reference to evaluate speaker similarity. Part3 used A/B test format to examine whether perceptually the proposed GuidedDiff improves articulation over the DuriTaco baseline. We skipped the comparison between the proposed GuidedDiff and the VC baseline because results from objective test clearly show that the VC baseline produces articulation nearly as good as natural speech. For the A/B test 15 sentences were selected from a script written by the target speaker. These sentences contain at least one word that the target speaker found difficult to articulate. The listening test is no-paid. In order to constrain the test duration to be within 15 minutes, we chose to randomly expose 10 out of 30 questions to the listeners in the first two parts.

 We received 87 effective responses after filtering out listeners who failed to spot the hidden reference. For the MUSHRA-style tests we performed a two-sided Wilcoxon signed-rank test on all pairs of stimuli, and corrected with Holm-Bonferroni. Results in Figure \ref{fig:nat_mushra} reveal similar pattern as the objective test: GuidedDiff improves the overall quality over the DuriTaco baseline and the target speaker's original recordings(\(p<0.001\)), though not as good as the VC baseline. One interesting finding is that even the DuriTaco baseline produces perceptually better speech than those in original recordings, indicating that a shared phone embedding across speakers does help to alleviate the articulation disorder in the synthetic voice. Results in Figure \ref{fig:sim_mushra} demonstrate the advantage of GuidedDiff over the VC baseline in terms of preserving the target speaker's individuality, with the voice in GuidedDiff and DuriTaco not significantly different from that in original recordings (\(p>0.09\)), while the same conclusion does not apply to DuriTaco+VC (\(p<0.001\)). The preference test result given in Figure \ref{fig:preference} further demonstrates that GuidedDiff synthesizes better articulation than the DuriTaco baseline. Readers are encouraged to visit \url{https://diffcorrect.github.io/} to listen to audio samples.

\section{Conclusion and Discussion}
We introduced the use of classifier-guided diffusion models for the creation of personalized synthetic voices from post-glossectomy speech. Experimental results on a real patient case show that the synthetic voice can restore articulation for phones that the target speaker had difficulties to articulate, and at the same time maximally preserve the target speaker's individuality. The proposed model takes around 1.4 seconds to synthesize 30 Cantonese characters on a single 2080Ti GPU. In the future we will extend the current model to take speech as additional condition, aiming at correcting improper articulations in the recordings directly. In this way the modified recordings with correct articulation can be used to train other TTS models to enable faster inference speed.

\section{Acknowledgements}
We thank all the participants in the listening test, for their kindness and contributions. Our deepest gratitude belongs to the young lady, who generously agreed to reveal her synthetic voice on the demo page. The first author is supported by the Hong Kong Ph.D. Fellowship Scheme of the Hong Kong Research Grants Council.

\bibliographystyle{IEEEtran}
\bibliography{mybib}

\end{document}